\documentclass{article}

\usepackage{arxiv}

\usepackage[utf8]{inputenc} 
\usepackage[T1]{fontenc}    
\usepackage[super,sort&compress,comma]{natbib} 
\usepackage{amsmath}
\usepackage[version=4]{mhchem}
\usepackage{graphicx}
\usepackage{url}            
\usepackage{booktabs}       
\usepackage{amsfonts}       
\usepackage{nicefrac}       
\usepackage{microtype}      
\usepackage{lipsum}
\usepackage{amssymb}
\usepackage{bm}
\usepackage{gensymb}
\usepackage[format=plain,justification=justified,singlelinecheck=false,font={stretch=1.125,small},labelfont=bf,labelsep=space]{caption}
\usepackage{float}
\usepackage{dcolumn}
\usepackage{booktabs}
\usepackage{dsfont}
\usepackage{overpic}
\usepackage{hyperref}
\hypersetup{
    colorlinks=blue,
    linkcolor=blue,
    filecolor=blue,      
    urlcolor=blue,
    citecolor=green
}

\title{On the possibility of chiral symmetry breaking in liquid hydrogen peroxide}

\author{
  Roberto Menta \\
  NEST, Scuola Normale Superiore, Piazza dei Cavalieri 7, I-56127 Pisa, Italy \\
  Department of Chemistry, Princeton University, Princeton, NJ 08544, USA \\
  \texttt{roberto.menta@sns.it}
  \And
  Pablo G. Debenedetti \\
  Department of Chemical and Biological Engineering, Princeton University, Princeton, NJ 08544, USA
  \And
  Roberto Car \\
  Department of Chemistry, Princeton University, Princeton, NJ 08544, USA
  \And
  Pablo M. Piaggi \\
  CIC nanoGUNE BRTA, Tolosa Hiribidea 76, 20018 Donostia-San Sebastián, Spain \\
  Ikerbasque, Basque Foundation for Science, 48013 Bilbao, Spain \\
  \texttt{pm.piaggi@nanogune.edu}
 }

\begin{document}
\maketitle

\begin{abstract}
Molecular chirality is a key concept in chemistry with implications for the origin of life and the manufacturing of pharmaceuticals.
Previous simulations of a chiral molecular model with an energetic bias towards homochiral interactions show a spontaneous symmetry-breaking transition from a supercritical racemic liquid into a subcritical liquid enriched in one of the two enantiomers.
Here, we employ molecular dynamics simulations in order to test the possible existence of this phenomenon in hydrogen peroxide, the smallest chiral molecule.
For this purpose, we study the fluid phase of this substance between 100 K and 1500 K, and from $10^{-4}$ GPa to 1 GPa.
We find a glass transition and we suggest that hydrogen bonds play a central role in such behavior.
We also test the possibility of observing chiral symmetry breaking by performing both constant temperature and cooling simulations at multiple pressures, and we do not observe the phenomenon.
An analysis of the structure of the liquid shows negligible differences between homochiral and heterochiral interactions, supporting the difficulty in observing chiral symmetry breaking.
If hydrogen peroxide manifests spontaneous chiral symmetry breaking, it likely takes place significantly below room temperature and is hidden by other phenomena, such as the glass transition or crystallization. More broadly, our results, and recent experimental observations, suggest that greater molecular complexity is needed for spontaneous chiral symmetry breaking in the liquid phase to occur.
\end{abstract}

\section*{Introduction}

Molecular chirality is a phenomenon characterized by the existence of two isomers, called enantiomers, which are non-superimposable mirror images of each other.
Understanding molecular chirality is highly relevant both for pharmaceutical science and biochemistry.
In pharmaceuticals, the various enantiomers of a drug can have dramatically different effects on the body, influencing the efficacy and safety of medical treatments\cite{li2018pharmaceutical,vargesson2015thalidomide}.
Moreover, in biochemistry, the building blocks of molecules synthesized by living organisms exist almost exclusively as a single enantiomer, for instance, naturally occurring amino acids are L-enantiomers while sugars are D-enantiomers.
The origin of this remarkable phenomenon, known as biological homochirality, is not yet fully understood\cite{blackmond2010origin}, and is an active area of research\cite{Ozturk2022, Ozturk2023}.
Hypotheses for its origin usually require a symmetry breaking agent and an amplification mechanism, but recent experiments have shown that spontaneous chiral symmetry breaking into isotropic liquids of opposite chirality can occur entirely in the liquid phase\cite{dressel2014chiral}.

We recently showed\cite{piaggi2023critical} that spontaneous breaking of chiral symmetry can occur in molecular dynamics (MD) simulations of a simple tetramer model\cite{latinwo2016molecular,Petsev2021} with a geometry inspired on hydrogen peroxide (\ce{H_2O_2}), the smallest chiral molecule.
When homochiral interactions are favored over heterochiral interactions, the liquid phase of this molecular model exhibits both ordinary vapor-liquid critical behavior and chirality-driven spontaneous symmetry breaking\cite{uralcan2021interconversion, wang2022fluid, Longo2023, piaggi2023critical}.
Furthermore, below the critical temperature for chiral symmetry breaking the liquid transforms into a phase enriched in one of the two enantiomers.
This model however was not designed nor parameterized to represent realistic intermolecular forces in \ce{H2O2} and it also ignored hydrogen-bonding interactions.

To see if spontaneous symmetry breaking can still occur in a more realistic model, here we study using MD simulations a model for \ce{H_2O_2} based on the standard OPLS force field parametrization\cite{jorgensen1996development}, but we modify the intramolecular dihedral angle interaction in order to agree approximately with experimental equilibrium dihedral angles and barriers for enantiomer interconversion.
We compute the properties of the liquid at ambient conditions and explore the fluid phase diagram from 100 K to 1500 K, and from $10^{-4}$ GPa to 1 GPa.
We find that this force field can describe qualitatively and even semi-quantitatively a number of properties of the gas and liquid phases, without any fine tuning.
We also investigate the possible spontaneous chiral symmetry breaking, but we do not find signatures of this phenomenon.
To understand this result, we analyze the interactions between enantiomers in the liquid phase and find negligible differences between the strength of homochiral and heterochiral interactions.
Finally, we characterize the glass transition in the liquid phase and find that it occurs at temperatures higher than those expected for spontaneous symmetry breaking in this system.

\section*{\label{sec:methods}Methods}
\subsection*{Model for \ce{H2O2}}
We consider a system of $N$ \ce{H2O2} molecules in a simulation box with periodic boundary conditions.
To describe the interactions between atoms in this system we draw from the OPLS force field\cite{jorgensen1996development} and define a potential energy surface, 
\begin{equation}
    U(\bm{R}) = U_{\text{intra}}(\bm{R}) + U_{\text{inter}}(\bm{R}) \ ,
\end{equation}
where $U_{\text{intra}}(\bm{R})$ and $U_{\text{inter}}(\bm{R})$ are the intra- and inter-molecular interactions, respectively, and $\bm{R} \in \mathbb{R}^{3 \times 4\times N}$ are the atomic coordinates.
The intramolecular interactions are given by
\begin{equation}\label{Uintra}
    \begin{array}{rlc}
         U_{\text{intra}}(\bm{R}) &= U_{\text{bonds}}(\bm{R}) + U_{\text{angles}}(\bm{R}) + U_{\text{dihedrals}}(\bm{R}) \ ,
    \end{array}
\end{equation}
with
\begin{equation}
    U_{\text{bonds}}(\bm{R}) = \sum_{i=1}^{N} \Big(\sum_{j=1}^3 K_{r,j} (\vert \bm{r}^i_j - \bm{r}^i_{j+1} \vert - r_{0,j})^2 \Big) \ ,
\end{equation}
where $\bm{r}^i_j$ is the position of the atom $j$ ($j=1,2,3,4)$ for each molecule $i$.
Then, $K_{r,j}$ and $r_{0,j}$ are the bond force constant and the equilibrium bond distance, respectively, for the bond between atoms $j$ and $j+1$.
Note that the force constant $K_{r,j}$ and all other force constants defined below include the usual 1/2 factor in harmonic potentials.
For \ce{H2O2} molecules, we have two bond types \ce{H}-\ce{O} and \ce{O}-\ce{O} with bond force constants $K_{r,\rm{OO}}$ and $K_{r,\rm{HO}}$, and equilibrium bond distances $r_{0,\rm{OO}}$ and $r_{0,\rm{HO}}$.
We have chosen such parameters based on the OPLS force field and we show their values in Table~\ref{tab:coefficients}.

\begin{table}[t!]
    \centering
    \begin{tabular}{cccccc}
    \toprule
    $K_{r,\rm{OO}} \Big[\frac{\rm{kcal}}{\rm{mol} \cdot \text{\AA}^2}\Big]$ & $K_{r,\rm{HO}} \Big[\frac{\rm{kcal}}{\rm{mol} \cdot \text{\AA}^2}\Big]$ & $r_{0,\rm{OO}}$ [\AA] & $r_{0,\rm{HO}}$ [\AA] & $K_{\theta} \Big[\frac{\rm{kcal}}{\rm{mol}}\Big]$  & $\theta_{0} \Big[  ^{\circ}  \Big]$ \\
    \midrule
    542.59 & 553 & 1.28 & 0.945 & 46.65 & 110.93 \\
    \bottomrule
    \end{tabular}

    \vspace{0.5cm} 

    \begin{tabular}{cccccccc}
    \toprule
    $V_{\ell=1,2,3,4} \Big[\frac{\rm{kcal}}{\rm{mol}}\Big]$ & $\epsilon \Big[\frac{\rm{kcal}}{\rm{mol}}\Big]$ & $\sigma$ [\AA] & $r_c$ [\AA] & $q_{\rm O}$ & $q_{\rm H}$ \\
    \midrule
    8, $-5$, 0, 0 & 0.17 & 3.12 & 10 & $-0.428$ & 0.428 \\
    \bottomrule
    \end{tabular}
    \caption{Parameters for the OPLS-based model for \ce{H2O2}. Energies are given in kcal/mol to facilitate reproducing the calculations using the software LAMMPS. Charges are expressed in units of the electron elementary charge.}
    \label{tab:coefficients}
\end{table}

The next term in Eq.~\eqref{Uintra} is $U_{\text{angles}}$, the energy due to the two HOO angles of each molecule, which is given by
\begin{equation}
    U_{\text{angles}}(\bm{R}) = \sum_{i=1}^{N} \Big(\sum_{k=1}^2 K_{\theta} (\theta^i_k - \theta_{0})^2 \Big) \ ,
\end{equation}
where $\theta^i_k$ with $k=1,2$ are the two bond angles in molecule $i$, $K_{\theta}$ and $\theta_{0}$ are the angle force constant and the equilibrium angle value, respectively.
We have also chosen $K_{\theta}$ and $\theta_{0}$ based on the OPLS force field (see Table~\ref{tab:coefficients}).

The last term in Eq.~\eqref{Uintra} is given by
\begin{equation}
     U_{\text{dihedrals}}(\bm{R}) 
     = \displaystyle \sum_{i=1}^{N} \Big(\sum_{\ell=1}^4 \dfrac{V_\ell}{2}[1 + (-1)^{\ell+1}\cos(\ell\phi^i)]\Big) \ ,
\end{equation}
where $\phi^i$ is the dihedral angle of molecule $i$. $\{V_{\ell}\}$ with $\ell=1,2,3,4$ are the energy coefficients for the dihedral interaction.
We recall that the hydrogen peroxide molecule exhibits an equilibrium dihedral angle of about $111.5^{\circ}$ and $93^{\circ}$ in the gas and solid (crystalline) phases, respectively\cite{giguere1950infra, hunt1965internal, giguere1983molecular}.
In the liquid phase, due to the reduced freedom of movement, the dihedral angle can vary but it typically remains close to the gas phase value.
However, we found that the standard parameterization of the OPLS force field results in non-chiral behavior with equilibrium dihedral angle $\sim 0^{\circ}$. 
We thus reparametrized the coefficients $V_{\ell}$ in order to obtain chiral conformers with equilibrium dihedral angle $\sim \pm 100^{\circ}$, and appropriate energy barriers for enantiomeric conversion through the cis and trans pathways\cite{hunt1965internal}.
The chosen coefficients $V_{\ell}$ are shown in Table~\ref{tab:coefficients}.

The inter-molecular interactions in our model are given by the standard 12/6 Lennard-Jones (LJ) potential for OO interactions and the Coulomb potential, i.e.,
\begin{equation}\label{LJ}
    \begin{split}
    &U_{\text{inter}}(\bm{R}) = \displaystyle\sum_{i=1}^{N}\sum_{j=1}^{N} \sum_{n=1}^4 \sum_{m=1}^4 f^{ij}_{nm} \Bigg\{4\epsilon\Bigg[\Bigg(\dfrac{\sigma}{\vert\bm{r}^i_n - \bm{r}^j_m\vert}\Bigg)^{12} - \\ \\ & \Bigg(\dfrac{\sigma}{\vert\bm{r}^i_n - \bm{r}^j_m\vert}\Bigg)^6 \ \Bigg] \cdot (1-\delta_{ij}) \Delta_{nm} +  \displaystyle \dfrac{q^i_n q^j_m}{4 \pi\epsilon_0 \vert\bm{r}^i_n - \bm{r}^j_m\vert^2}\Bigg\}\ .
    \end{split}
\end{equation}
In Eq.~\eqref{LJ}, $\sigma$ and $\epsilon$ are the LJ parameters, $\epsilon_0$ is the permittivity of vacuum, and $q^i_n$ and $q^j_m$ are the charges on atoms $i$ and $j$ in molecules $n$ and $m$, respectively.
$q^i_n$ and $q^j_m$ can take two values: $q_{\rm O}$ and $q_{\rm H}$ according to the chemical species of each atom.
Furthermore, in Eq.~\eqref{LJ}, $\delta_{ij}$ is the usual Kronecker's delta while $\Delta_{nm}$ is a two-index tensor which accounts for the fact that the LJ potential acts only on non-bonded OO interactions. The latter can be written explicitly as: $\Delta_{nm} = \delta_{2m}\delta_{n2} + \delta_{3m}\delta_{n3} + \delta_{2m}\delta_{n3} + \delta_{3m}\delta_{n2}$. Furthermore, $f^{ij}_{nm}$ is the so-called ``fudge factor'' and has the following properties: a) if $n=m$ then $f^{14}_{nn} = 0.5$, otherwise $f^{ij}_{nn} = 0$; b) if $n\neq m$ then $f^{ij}_{nm} = 1, \ \forall \ i,j$, viz.~full interaction. The parameters of the non-bonded interactions $\epsilon$, $\sigma$, $q_{\rm O}$, $q_{\rm H}$, and $r_c$ for the OPLS molecular model are reported in Table~\ref{tab:coefficients}.
We note that in our simulations the Lennard-Jones interaction was truncated at the cutoff $r_c$, while the Coulomb interactions are computed in real space up to $r_c$ and the Ewald summation\cite{TOUKMAJI199673} is used to take into account long-range interactions (see Methods section for further details).
The masses of the two species of atoms were set to $m_{\rm O} = 15.999 \ \rm{g/mol}$, and $m_{\rm H} = 2 \ \rm{g/mol}$.
Note that we have used the mass for deuterium, instead of hydrogen, to allow for a longer timestep in the molecular dynamics simulations.

\subsection*{Molecular dynamics simulations}
We performed all simulations using the MD engine LAMMPS\cite{thompson2022lammps} and a time step of 0.5 fs.
Note that our model has flexible bonds and thus requires a shorter timestep compared to rigid models.
We set the temperature using the stochastic velocity rescaling thermostat of Bussi et al.\cite{bussi2007canonical} with a relaxation time of 0.1 ps, and we controlled the pressure with an istotropic Parrinello-Rahman barostat \cite{parrinello1981polymorphic, martyna1994constant} with a relaxation time of 1 ps.
Long range Coulomb interactions beyond $r_c$ were computed using the particle-particle particle-mesh (P$^3$M) method\cite{Hockney2021-bo} with a target relative error in the forces below $10^{-5}$.

\paragraph{Calculation of liquid-glass transition temperature.}
We computed the glass transition temperature $T_g$ for different pressures in the range $10^{-4}$ GPa to 1 GPa using the following method.
For each pressure, we performed cooling simulations from an initial temperature of $400$ K to a final temperature of $100$ K in 500 ns.
We then calculated the mean enthalpy in windows of 150 ps.
As we shall see, the mean enthalpy shows a clear change in slope as the temperature is lowered, which is a signature of the glass transition.
We thus defined two temperature regimes above and below the change in slope, and we fit straight lines to the enthalpy vs temperature data in each temperature regime.
Finally, we defined $T_g$ as the temperature of intersection between the two straight lines.
These calculations were performed using $N=512$ molecules, and we repeated the simulations at each pressure 40 times, each with a different initial velocity distribution, to improve the accuracy in the estimation of the glass transition temperature.

\begin{figure}[H]
    \centering
    \includegraphics[width=1.0\columnwidth]{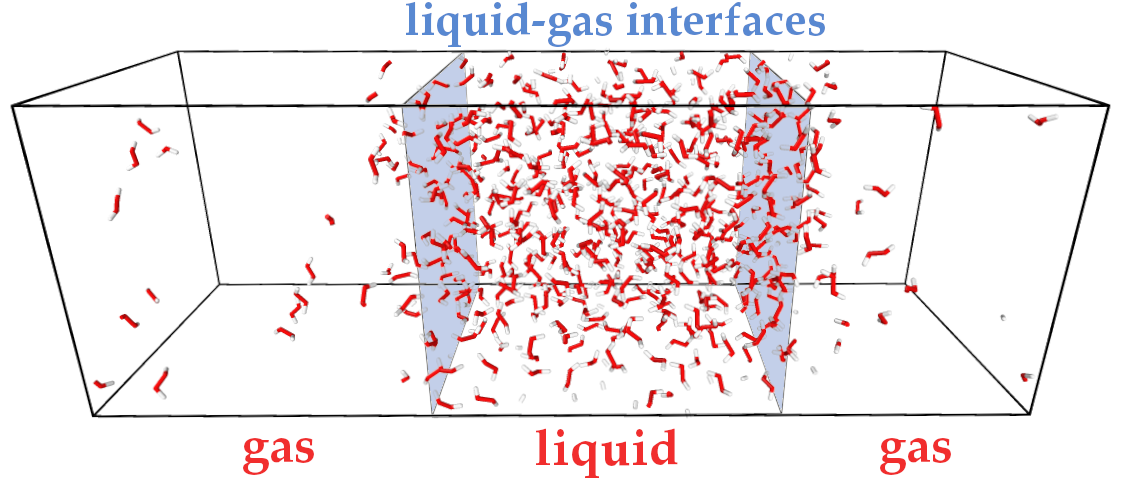}
    \caption{Snapshot of liquid-gas coexistence configuration at $T=500$ K. The flat liquid-gas interfaces are depicted schematically.}
    \label{fig:liquid-gas500K}
\end{figure}

\paragraph{Calculation of liquid-gas coexistence curve.}
To compute the coexistence curve between the liquid and gas phases, we used a simulation box with dimensions $27 \ \text{\AA} \times 27 \ \text{\AA} \times 81 \ \text{\AA}$.
Note that one dimension, the $z$-direction, is much larger than the other two.
Initially, we placed a liquid slab at the center of the simulation box, such that there were two flat liquid-vacuum interfaces.
Then, we equilibrated the system at constant temperature and with fixed box dimensions, and we computed the normal pressure along the $z$-direction.
During the equilibration some molecules leave the liquid phase and go into the gas phase.
Once that the simulation reaches equilibrium, the average pressure of the system along $z$ is the liquid-gas coexistence pressure at the chosen temperature.
We repeated this procedure for multiple temperatures, namely $T=400,420,450,480,500,520$ and 550 K, to determine that coexistence curve ($P$ vs. $T$).
A snapshot of a typical liquid-gas coexistence configuration at 500 K is displayed in Fig.~\ref{fig:liquid-gas500K}.

\section*{Results}

Before describing the results of the fluid phase diagram and the possible chiral symmetry breaking, we show some properties of \ce{H2O2} in the gas and the liquid phases within the description of our force field.

\subsection*{Properties of the molecule in the gas phase}

\begin{table}[b]
    \centering
    \begin{tabular}{@{}ccc@{}}
    \toprule
    \textbf{Parameters} & \textbf{This work} & \textbf{Experiment}\cite{hunt1965internal} \\
    \midrule
    $r_{0,\rm{OO}}$ & 1.28 \AA & 1.475 \AA \\
    $r_{r,\rm{HO}}$ & 0.945 \AA & 0.95 \AA \\
    $\theta_0$ & $110.93^{\circ}$ & $94.8^{\circ}$ \\
    $\phi_0$ & $108^{\circ}$ & $111.5^{\circ}$ \\
    
    \bottomrule
    \end{tabular}
    \caption{Comparison of structural properties of our model at 0 K with experimental data\cite{hunt1965internal}. $r_{0,\rm{OO}}$ and $r_{0,\rm{HO}}$ are the equilibrium OO and OH distances in the molecule, $\theta_{0}$ is the HOO angle, and $\phi_0$ is the HOOH dihedral angle.}
    \label{tab:experiment_gas_phase}
\end{table}

We first computed the geometry of the \ce{H2O2} molecule at 0 K.
In Table~\ref{tab:experiment_gas_phase} we show the equilibrium OO and OH distances in the molecule ($r_{0,\rm{OO}}$ and $r_{r,\rm{HO}}$), the HOO angle ($\theta_{0}$), and the HOOH dihedral angle ($\phi_0$).
In the same table we compare our results with experimental data at finite $T$\cite{hunt1965internal}.
Overall, the agreement between the molecular geometry in our model and in the experiment is good, except for the $r_{0,\rm{OO}}$ distance which is underestimated by around 20\%.
This is a result of the OPLS standard parameterization, and could be improved with a new choice of parameters.
The most important property, i.e.~the equilibrium dihedral angle $\phi_0$ that characterizes the molecule's chirality, is in good agreement with the experiment\cite{hunt1965internal}.

\begin{figure}[t]
    \centering
    \includegraphics[width=0.7\columnwidth]{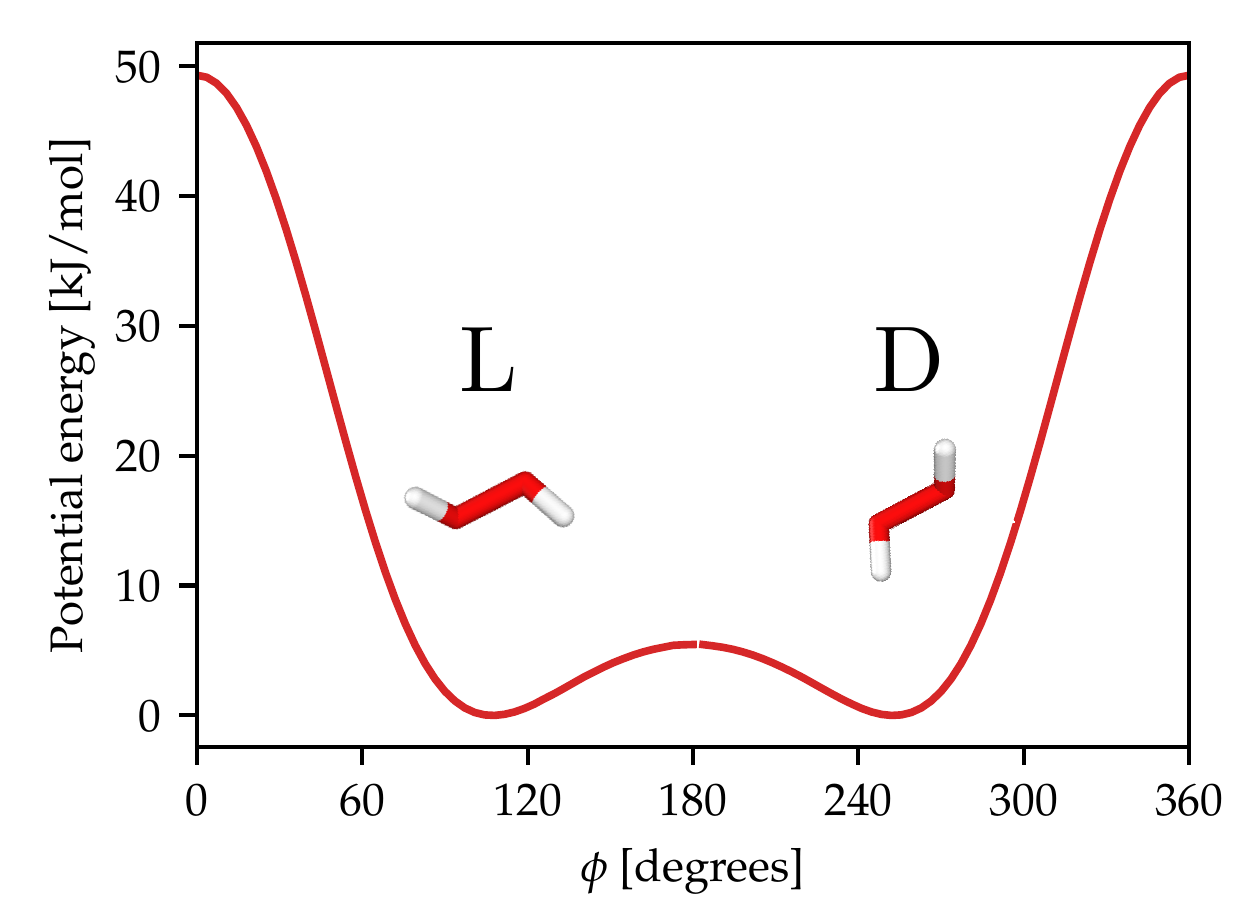}
    \caption{Potential energy vs. dihedral angle $\phi$ for a single \ce{H2O2} molecule in the gas phase at $0$ K.}
    \label{fig:energy-vs-dihedral}
\end{figure}

In Fig.~\ref{fig:energy-vs-dihedral} we show the total potential energy as a function of the dihedral angle $\phi$.
The curve shows two symmetric minima at around $\phi\sim \pm108^{\circ}$ and two different barriers for the trans and cis pathways, with energy barriers of 5 and 49 kJ/mol, respectively.
These are in satisfactory agreement with experiments that report energy barriers of 5 and 30 kJ/mol for the trans and cis pathways, respectively\cite{hunt1965internal}.

\subsection*{Properties of the liquid at 300 K and 1 atm}

\begin{figure}[t]
    \centering
    \includegraphics[width=0.7\columnwidth]{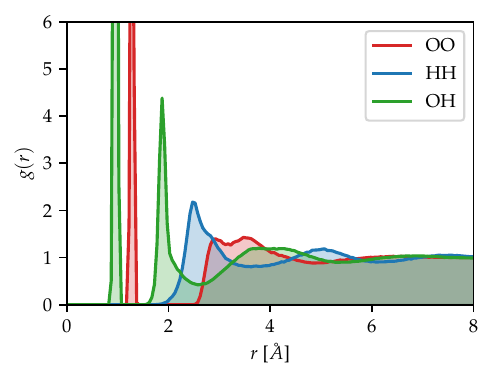}
    \caption{Radial distribution function between oxygen atoms (OO), hydrogen atoms (HH) and oxygen-hydrogen pairs (OH) calculated using 512 \ce{H2O2} molecules in the liquid phase at 300 K and 1 atm. }
    \label{fig:rdf-liquid}
\end{figure}

We now turn our attention to the properties of the liquid phase at 300 K and 1 atm.
In Fig.~\ref{fig:rdf-liquid} we show the OO, OH, and HH radial distribution functions (RDFs) for liquid \ce{H2O2}.
The OO and OH RDFs show narrow and high peaks at short distances, namely 1.28 \AA\ and 0.945 \AA, corresponding to the intramolecular covalent bonds already described for the molecule in the gas phase.
The OH RDF shows a peak at around 2 \AA, signaling the formation of hydrogen bonds in the liquid phase.
In Ref.~\citenum{MARTINSCOSTA2007341}, a similar behavior was found for a single \ce{H2O2} molecule in aqueous solution using a combined quantum/classical force field.
The OO RDF shows a broad feature associated with neighboring molecules with distances in the range 2.8 -- 3.9 \AA\ .
Beyond 6 \AA\ all RDFs show negligible structure.
Experimental information on the molecular structure of the liquid is very limited\cite{Randall1937-gx} and for this reason we compare our results with the experimental crystal structure of \ce{H2O2}\cite{abrahams1951crystal}.
These experiments show that the closest OO intermolecular distances are 2.8 -- 2.9 \AA\ and 3.4 -- 3.6 \AA\, which are the distances between the O atoms belonging to neighboring molecules.
Thus, the broad feature found in our calculations for the OO RDF is associated to OO distances between neighboring molecules, and this is consistent with the experimental crystalline structure of \ce{H2O2}\cite{abrahams1951crystal}.

\begin{figure}[H]
    \centering    \includegraphics[width=0.7\columnwidth]{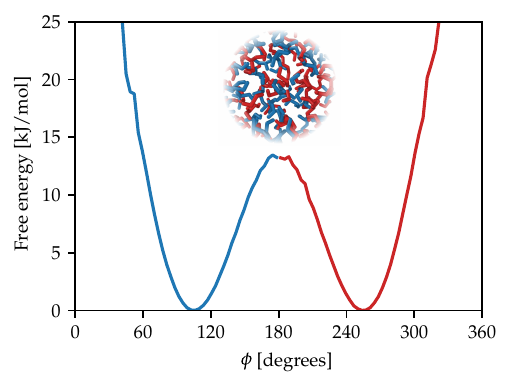}
    \caption{Free energy vs.~dihedral angle $\phi$ for 512 \ce{H2O2} molecules in the liquid phase at 300 K and 1 atm. The snapshot of the trajectory shows L and D enantiomers colored in blue and red, respectively. The curve has been colored accordingly.}
    \label{fig:freeenergy-vs-dihedral}
\end{figure}

We also computed the free energy as a function of the average dihedral angle taken over all molecules in the system $F(\langle \phi \rangle)$, which is shown in Fig.~\ref{fig:freeenergy-vs-dihedral}.
$F(\langle \phi \rangle)$ is calculated from the histogram of the average dihedral angle $P(\langle \phi \rangle)$ using $F(\langle \phi \rangle)=-\beta^{-1} \log[P(\langle \phi \rangle)]$, with $\beta=1/k_BT$ where $k_B$ is the Boltzmann's constant.
The curve in Fig.~\ref{fig:freeenergy-vs-dihedral} shows that the chiral behavior is maintained in the liquid phase, with similar equilibrium angles, and that the energy barrier for the transformation along the trans configuration is higher than in the gas phase.
This increase in the barrier may have its origin in the steric hindrance for the rotation of the dihedral angle caused by the surrounding molecules in the liquid.

We found that the density of the liquid at these conditions is $1.31$ g/cm$^3$, which is somewhat lower that the experimental density $1.44$ g/cm$^3$ at the same conditions.

\begin{figure*}[!t]
\includegraphics[width=1.0\columnwidth]{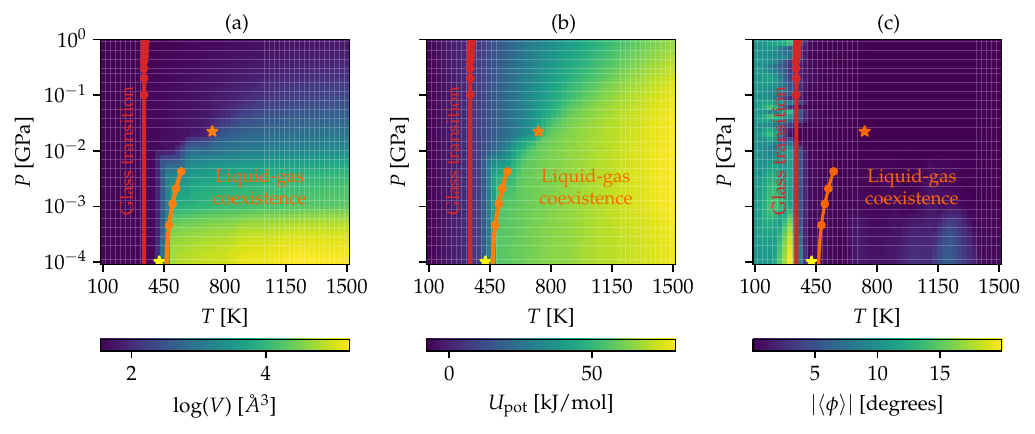}
\caption{Color plots in pressure vs. temperature for (a) the volume per molecule, (b) the potential energy, and (c) the mean dihedral angle. The glass transition and liquid-gas coexistence curves are shown in red and orange, respectively. The topmost data point in the liquid-gas coexistence curve is the highest temperature for which liquid-gas coexistence was observed. The orange and yellow stars in the figures correspond to the experimental critical and atmospheric pressure boiling points, respectively\cite{NIKITIN1995945}.}
\label{fig:phase-diagram}
\end{figure*}

\subsection*{Fluid phase diagram}

In order to understand the phase behavior and thermodynamics of the liquid and vapor, we performed simulations at constant temperatures and pressures in the range 100 K to 1500 K and from $10^{-4}$ GPa to $1$ GPa with equally spaced temperatures with a step of 100 K and equally spaced pressures in a log scale, such that there were 10 different pressures per decade.
From these simulations we computed the average volume per molecule, the average potential energy (intra and intermolecular), and the average dihedral angle, which are shown as a function of temperature and pressure in Fig.~\ref{fig:phase-diagram}(a), (b), and (c), respectively.
The average volume, shown in Fig.~\ref{fig:phase-diagram}(a), changes rapidly at around 500 K and below $10^{-2}$ GPa -- a clear signature of the liquid to vapor transition.
Instead, the change from low to high volumes is smoother above $10^{-2}$ GPa, which is compatible with the behavior expected in the supercritical fluid.
A similar behavior is observed for the potential energy in Fig.~\ref{fig:phase-diagram}(b).
The average dihedral angle, shown in Fig.~\ref{fig:phase-diagram}(c), is close to zero for temperatures above 300 K and for any pressure, yet it becomes different from zero below this temperature.
Note that even though each molecule is chiral, with a dihedral angle of around 100$^\circ$, above 300 K there is a statistically equal number of D and L molecules in the system such that the dihedral angle averaged over all molecules in the system is close to zero, i.e., a racemic mixture.
Instead, below 300 K the average dihedral is different from zero, which is related to an unequal number of D and L molecules.
This behavior could be a consequence of the symmetry breaking described in Refs.~\citenum{wang2022fluid} and~\citenum{piaggi2023critical} for a chiral molecular model, or it could also be a consequence of a lack of ergodicity at low temperatures connected to a glass transition.

\begin{figure}[!t]
    \centering
    \includegraphics[width=0.8\columnwidth]{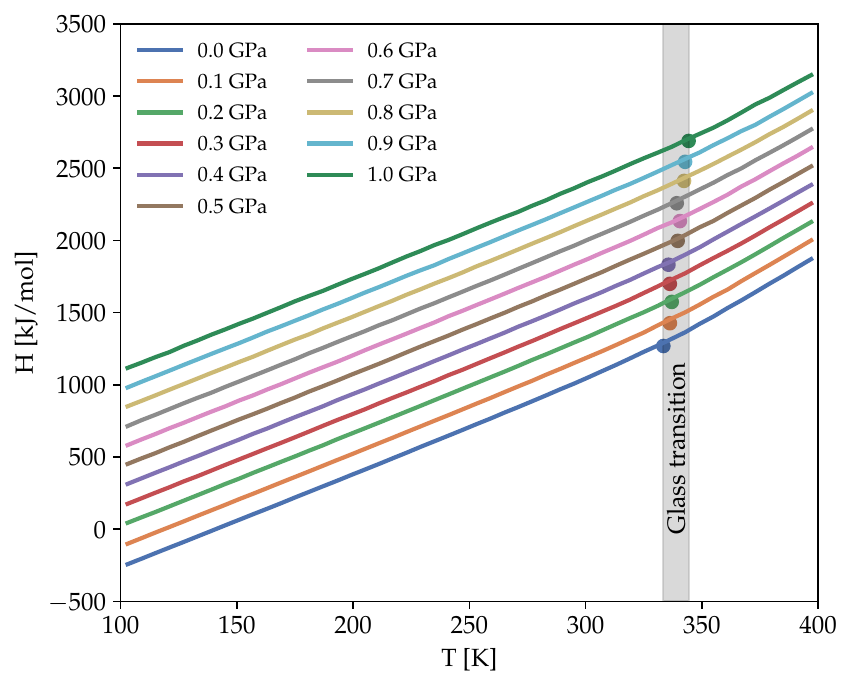}
    \caption{Enthalpy vs temperature curves obtained from cooling simulations from 400 K to 100 K in 500 ns at fixed pressures. Curves correspond to the enthalpy and temperature averaged over windows of 150 ps. The glass transition temperatures are shown using circles with colors matching the curves for the corresponding pressures. The error in the glass transition temperature for each point is estimated to be $\pm 4$ K, using the standard deviation of the enthalpy based on 40 independent cooling simulations at each pressure. The temperature window where the glass transition takes place is marked in gray to guide the eye.}
    \label{fig:glass-transition}
\end{figure}

To shed light on this behavior, we computed the glass transition curve using the technique described in the Methods section.
In Fig.~\ref{fig:glass-transition} we show the enthalpy vs temperature for different pressures between $10^{-4}$ GPa to $1$ GPa.
At all pressures, we see a change of slope in the enthalpy vs temperature curve which signals the glass transition.
The glass transition temperature at atmospheric pressure is located at around 340 K and increases slightly with pressure.
In the Supporting Information we analyze the behavior of the cumulative mean square displacement to verify our estimate of the glass transition temperature and find satisfactory agreement.
For comparison with previous calculations, we show the glass transition temperature vs pressure in Fig.~\ref{fig:phase-diagram} (red data points and curve).
This curve is in good agreement with the temperature for which the dihedral angles no longer average to zero in our previous simulations.
Therefore, the behavior of the average dihedral angle could be explained by the lack of ergodicity below the glass transition temperature.

We also computed the liquid-gas coexistence curve following the procedure described in the Methods section.
These curves are shown in the plots in Fig.~\ref{fig:phase-diagram} together with experimental data for the boiling point at $10^{-4}$ GPa and the critical point.
The boiling point at $10^{-4}$ GPa in our model is $460 \pm 10$ K, which compares favorably with the experimental boiling point at 423.3 K\cite{NIKITIN1995945}.
From the coexistence curve, we also estimated a lower bound for the critical point in our model by taking the highest temperature at which there is still liquid-gas coexistence. 
The critical point thus calculated is located at around 550 K and $4 \times 10^{-3}$ GPa, which is underestimated compared to the experimental counterpart at 728 K and 0.02 GPa\cite{NIKITIN1995945}.
A more accurate estimate of the critical point could have been obtained by computing the density of the liquid and vapor phases as a function of temperature and fitting the resulting data (see for example Ref.~\citenum{panagiotopoulos2024sequence}).
The boiling and critical points are depicted as yellow and orange stars in Fig.~\ref{fig:phase-diagram}, respectively.
Comparing the liquid-gas coexistence curve with the experimental data, we observe that the trend is captured very well by our force field.

\subsection*{Possible chiral symmetry breaking}
The central aim of this work is to investigate the possibility of chiral symmetry breaking in liquid \ce{H_2O_2}. In the theory of symmetry breaking phase transitions\cite{Landau:1937obd}, the order parameter of the model represents an important tool to describe the critical behavior of a statistical system. In our case, such order parameter is given by the average dihedral angle $\langle\phi\rangle$. Its behavior as a function of temperature and pressure, already discussed in the previous section, is shown in Fig.~\ref{fig:phase-diagram}(c). The origin of the non-zero average dihedral angle below $\sim$ 340 K needs to be investigated in greater detail.

\begin{figure}[H]
    \centering   
    \includegraphics[width=\textwidth]{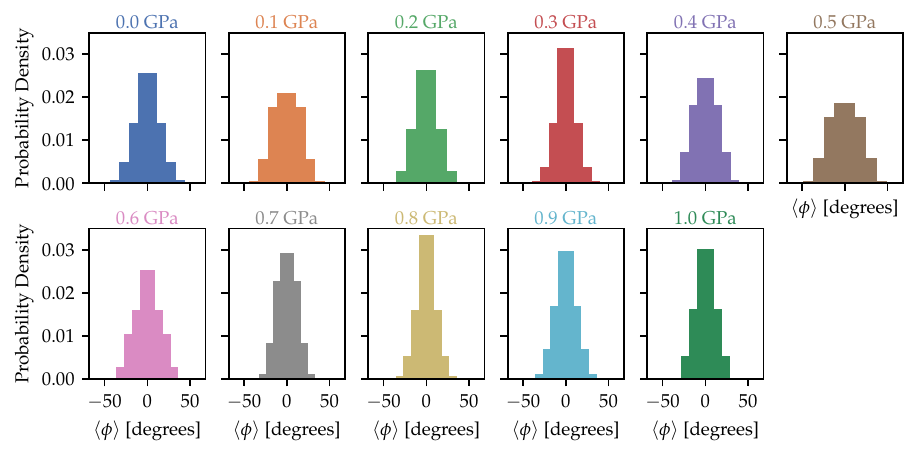}
    \caption{Probability density of the order parameter (average dihedral angle) calculated using simulations at constant temperature 340 K and multiple constant pressures in the range 0-1 GPa. The target pressure used in the calculation of each probability density is indicated at the top of each plot.}
    \label{fig:distr_CSB_340}
\end{figure}

In order to do that, we first simulated the liquid at 340 K and multiple pressures for 500 ns starting from a racemic mixture of L and D molecules.
340 K is the lowest temperature at which we are able to equilibrate the liquid in the standard time scales accessible to molecular dynamics simulations.
We note that 340 K matches approximately the glass transition temperature and is also the minimum temperature for which the order parameter $\langle\phi\rangle$ averages to $\sim 0$.
Thus, 340 K is the lowest temperature at which we are able to assess rigorously the possibility of spontaneous symmetry breaking.
In Fig.~\ref{fig:distr_CSB_340} we show probability densities of the order parameter during these simulations.
For all pressures we observe a single-peaked distribution centered at $\phi \approx 0$ compatible with the absence of symmetry breaking.

\begin{figure}[H]
    \centering
    \includegraphics[width=0.55\columnwidth]{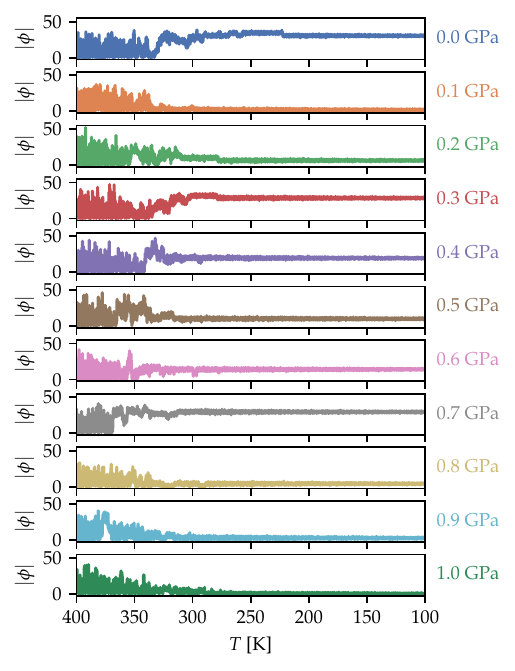}
    \caption{Cooling simulations from 400 K to 100 K in 500 ns at fixed pressures.}
    \label{fig:quenches} 
\end{figure}

To gather evidence about the possibility of symmetry breaking below the glass transition, we employed cooling simulations performed over 500 ns starting from 400 K and ending at 100 K at different fixed pressures from $10^{-4}$ GPa to $1$ GPa.
In Fig.~\ref{fig:quenches} we show the order parameter vs temperature during such simulations. We observe that below $\sim$ 340 K (i.e.~starting from around 150 ns) in some of the simulations the absolute value of the average dihedral angle $\vert \langle\phi\rangle \vert$ tends to non-zero values, up to $40^{\circ}$.

\begin{figure}[H]
    \centering   
    \includegraphics[width=\textwidth]{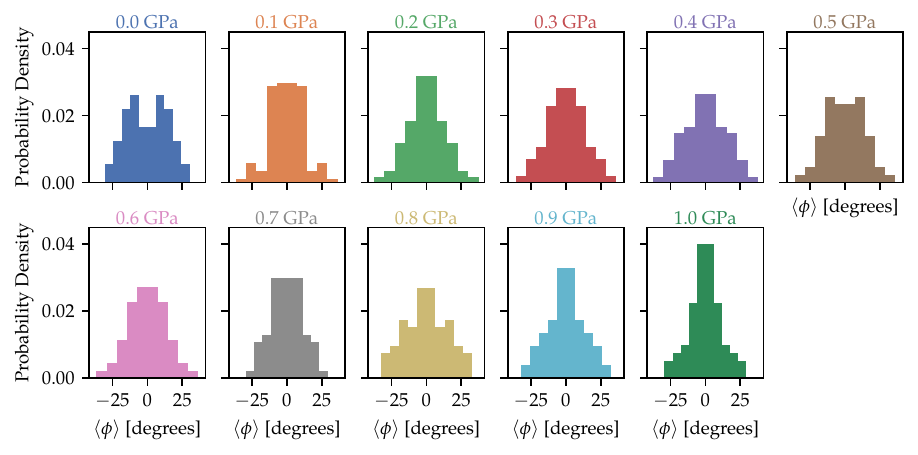}
    \caption{Probability densities of the final average dihedral angle in 40 
    cooling simulations at several fixed pressures. For pressures of $0.0, 0.1, 0.5$, and $0.6$ GPa, the normalized histograms are based on 60 simulations.}
    \label{fig:distr_CSB}
\end{figure}

In order to analyze the statistical uncertainty in the final values of $\langle\phi\rangle$, we repeated the cooling simulations 40 times for each pressure $P \in [10^{-4},1]$ GPa and we constructed the corresponding probability densities of the final values of $\langle\phi\rangle$. In Fig.~\ref{fig:distr_CSB}, we show probability densities at several fixed pressures. We observe that the majority of the normalized histograms show a single peak, signaling an absence of chiral symmetry breaking. A minority of the distributions are bimodal, yet the origin of such behavior can be traced to a lack of statistics rather than chiral symmetry breaking. Indeed, for these cases we checked that increasing the number of cooling simulations from 40 to 60 makes the distributions unimodal. 
For instance, probability densities at $0.0, 0.1, 0.5$, and $0.6$ GPa, when constructed from 40 cooling simulations, appeared as bimodal distributions.
By reconstructing these normalized histograms with a larger sample size, we confirmed that most of the distributions, as shown in Fig.~\ref{fig:distr_CSB}, exhibit a single peak.
This analysis suggests that our model of \ce{H_2O_2} does not show chiral symmetry breaking in the fluid phase. 
In the Supporting Information we report the results of constant temperature simulations below the glass transition, which also do not show spontaneous chiral symmetry breaking and agree with the conclusion reported here.

\begin{figure}[H]
    \centering   
    \includegraphics[width=\textwidth]{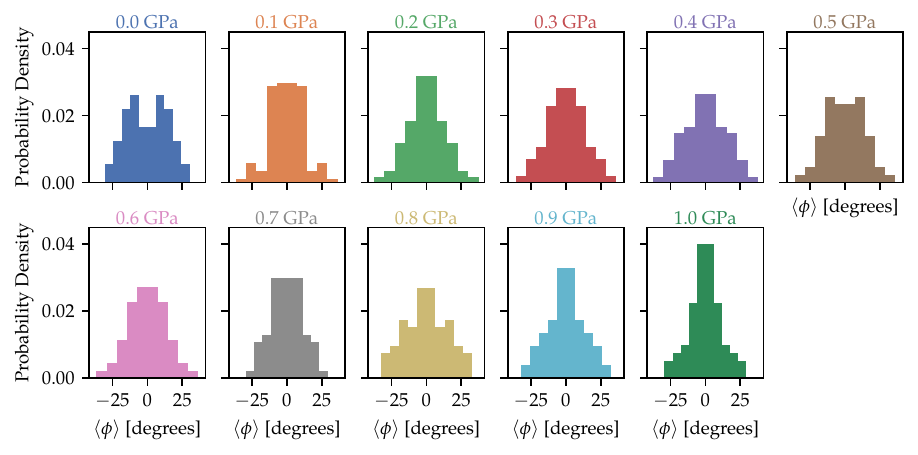}
    \caption{Probability densities of the final average dihedral angle in 40 
    cooling simulations at several fixed pressures. For pressures of $0.0, 0.1, 0.5$, and $0.6$ GPa, the normalized histograms are based on 60 simulations.}
    \label{fig:distr_CSB}
\end{figure}

\begin{figure}[t]
    \centering
    \includegraphics[width=0.8\columnwidth]{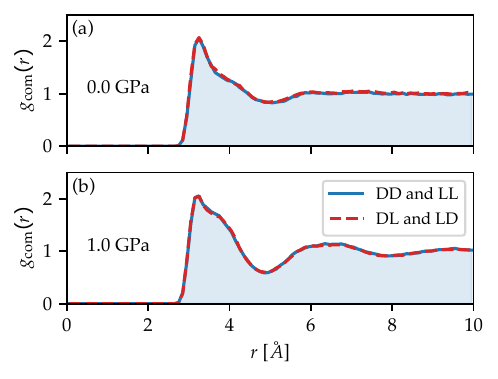}
    \caption{Radial distribution functions (RDFs) for DD, LL and DL (LD) interacting molecules at 300 K and two different pressures. Results at $\sim$ 0 GPa and 1 GPa are shown in panel (a) and (b), respectively.}
    \label{fig:rdf}
\end{figure}

The origin of the spontaneous symmetry breaking is the difference in the interaction strength between enantiomers of the same chirality (homochiral interactions), i.e.~DD or LL, and enantiomers of different chirality (heterochiral interactions), i.e.~DL or LD.
It is thus interesting to compute the radial distribution function (RDF) between enantiomers of the same chirality (DD and LL) and enantiomers of different chirality (DL and LD).
In Fig.~\ref{fig:rdf}(a), the RDFs for the molecular center of mass are shown at fixed pressure $P=10^{-4}$ GPa.
In that figure, we observe negligible differences between the RDF for pairs of enantiomers of the same chirality (DD and LL) and of different chirality (DL and LD).
This indicates a similar strength of homochiral and heterochiral interactions.
At higher pressures, the higher density of the liquid could favor homochiral interactions due to a more efficient packing of the \ce{H2O2} molecules.
Indeed, this was confirmed in Ref.~\citenum{uralcan2021interconversion} for the molecular model proposed by Latinwo et al.~\cite{latinwo2016molecular}.
In Fig.~\ref{fig:rdf}(b), we show the RDFs for the molecular center of mass at fixed pressure $P=1$ GPa.
Even at these higher pressures we see no evidence of homochiral interactions favored over heterochiral interactions.
This result is compatible with the absence of symmetry breaking found in our simulations.

\section*{Discussion \& Conclusions}
We have explored the possibility of observing chiral symmetry breaking in the liquid phase of \ce{H2O2} using molecular dynamics simulations based on a semi-empirical force field for this substance.
This symmetry breaking phenomenon has already been observed in simulations of a chiral tetramer with the same geometry as the \ce{H2O2} molecule\cite{latinwo2016molecular,piaggi2023critical}.
Here, we analyzed the fluid phase diagram of \ce{H2O2} and find that the liquid-gas phase boundary is reproduced reasonably well by our force field (see Fig.~\ref{fig:phase-diagram}).
Furthermore, we found evidence of a glass transition at around 340 K and the temperature of the glass transition increases slightly with pressure.
The glass transition, not reached in previous studies of the chiral tetramer model of Latinwo et al.~\cite{latinwo2016molecular}, arises as a consequence of the hydrogen bond interactions in our force field for \ce{H2O2}.
There is also experimental evidence for a glass transition in \ce{H2O2}\cite{ghormley1957warming,giguere1954hydrogen}, although the exact location of the glass transition temperature is unknown.
We suggest that new experiments could be performed in order to pinpoint precisely the glass transition temperature in this system.

We also evaluated the possibility of chiral symmetry breaking in the liquid by performing constant temperature as well cooling simulations at fixed pressure and found that, within our force field, \ce{H2O2} does not manifest this phenomenon.
We traced back this behavior to a negligible difference between the strength of homochiral and heterochiral interactions.
Such small difference between interaction strengths indicates that symmetry breaking could only take place well below room temperature.
Furthermore, the glass transition as well as crystallization at 272.72 K (experimental), might occur at higher temperatures than chiral symmetry breaking, thus hiding the phenomenon.
This is reminiscent of the situation in liquid water, where the glass transition and crystallization greatly hamper the study of the liquid-liquid phase transition\cite{Debendetti2003}.
While the force field employed here has limitations in representing the properties of \ce{H2O2} accurately, the much greater strength of H-bond interactions compared with the difference between homochiral and heterochiral interactions suggests that the results obtained here should persist even in more realistic models for this system.
Our study also indicates that more complex molecular systems, such as those studied by Dressel et al.\cite{dressel2014chiral}, may be necessary to observe the phenomenon of chiral symmetry breaking.
While our current work focuses on a specific system, future research could explore a broad range of chiral molecules to determine the existence of this phenomenon in other systems.

\subsection*{Acknowledgements}
This work was conducted within the center Chemistry in Solution and at Interfaces funded by the USA Department of Energy under Award DE-SC0019394. Simulations reported here were performed using the Princeton Research Computing resources at Princeton University which is consortium of groups including the Princeton Institute for Computational Science and Engineering and the Princeton University Office of Information Technology’s Research Computing department.
R.M. acknowledges funding from the Italian Scientists and Scholars in North America Foundation (ISSNAF) scholarship. R.M. thanks Enrica D'Ettorre, Stefano Bolognesi and the Physics Department ``E.~Fermi'' of the University of Pisa for facilitating his visit to Princeton. R.M. also expresses gratitude to the Frick Chemistry Department of Princeton University for their hospitality.
P.M.P. acknowledges funding from the Marie Skłodowska-Curie Cofund Programme of European Commission project H2020-MSCA-COFUND-2020-101034228-WOLFRAM2.

\subsection*{Data Availability}
Input and analysis files to reproduce the results reported here are available at \url{https://doi.org/10.5281/zenodo.15039101}.

\section*{Author Declarations}
\subsection*{Conflict of Interest}
The authors have no conflicts to disclose.
\subsection*{Author Contributions}
\textbf{Roberto Menta:} Investigation (co-lead); Writing original draft (co-lead); Conceptualization (supporting); Data curation (co-lead); Writing review \& editing (co-lead). \textbf{Pablo G. Debenedetti:} Conceptualization (supporting); Supervision (supporting); Writing – review \& editing (equal). \textbf{Roberto Car:} Conceptualization (supporting); Supervision (supporting); Writing – review \& editing (equal).
\textbf{Pablo M. Piaggi:} Investigation (co-lead); Writing original draft (co-lead); Conceptualization (lead); Supervision (lead); Data curation (co-lead); Writing review \& editing (co-lead).

\providecommand{\latin}[1]{#1}
\makeatletter
\providecommand{\doi}
  {\begingroup\let\do\@makeother\dospecials
  \catcode`\{=1 \catcode`\}=2 \doi@aux}
\providecommand{\doi@aux}[1]{\endgroup\texttt{#1}}
\makeatother
\providecommand*\mcitethebibliography{\thebibliography}
\csname @ifundefined\endcsname{endmcitethebibliography}  {\let\endmcitethebibliography\endthebibliography}{}

\clearpage
\setcounter{figure}{0}
\renewcommand{\thefigure}{S\arabic{figure}}

\section*{
\centering
\Large
Supporting Information to \\ ``On the possibility of chiral symmetry breaking in liquid hydrogen peroxide''}

In this Supporting Information we report the results of constant temperature simulations of the liquid phase of hydrogen peroxide, as described by our force field.
These results provide additional evidence for the absence of spontaneous chiral symmetry breaking in our model for hydrogen peroxide.

\section*{Additional simulations}

\begin{figure}[H]
    \centering   
    \includegraphics[width=\textwidth]{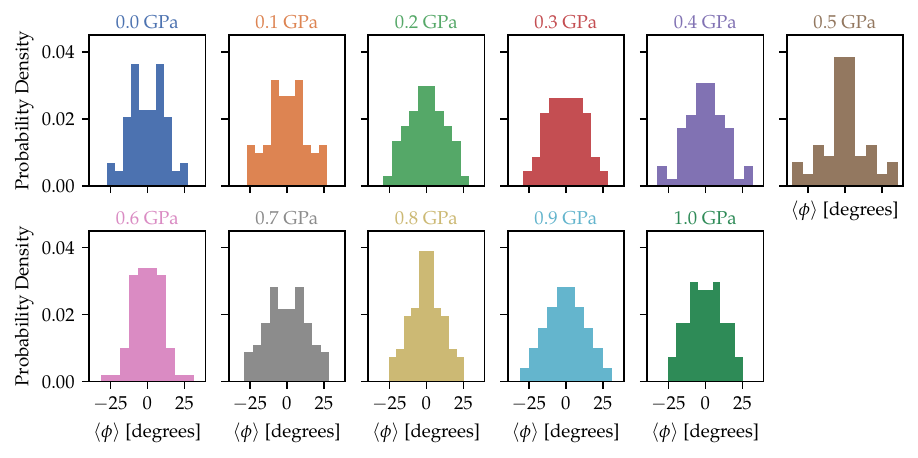}
    \caption{Probability densities of the final dihedral angle in 40 quench simulations with target temperature 230 K and at several fixed pressures. The probability densities are unimodal owing to the good statistics obtained from 40 independent simulations, even if the low temperature results in a liquid out of equilibrium.}
    \label{fig:histograms_230K}
\end{figure}

\begin{figure}[H]
    \centering   
    \includegraphics[width=\textwidth]{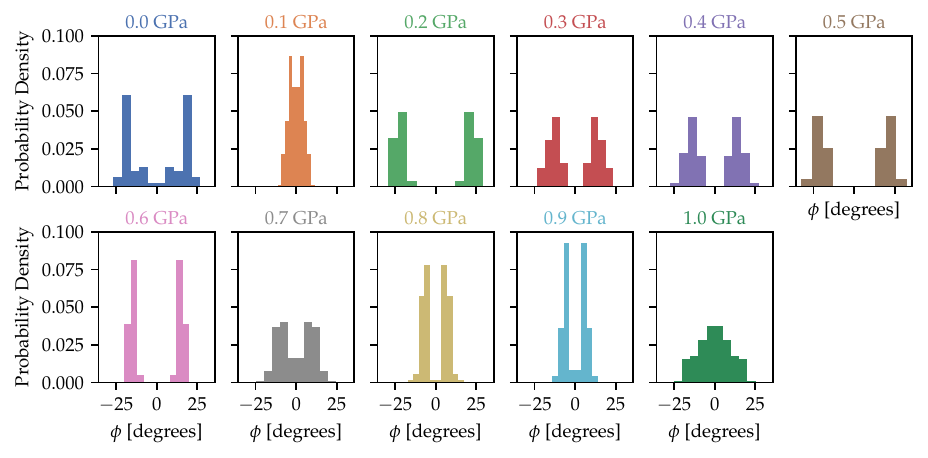}
    \caption{Probability densities of the dihedral angle in a single
    quench from with target temperature 280 K and at several fixed pressures. Here, a single quench simulations at each pressure is not sufficient to provide the statistics needed to show unimodal behavior.}
    \label{fig:histograms_280K}
\end{figure}

\begin{figure}[H]
    \centering   
    \includegraphics[width=\textwidth]{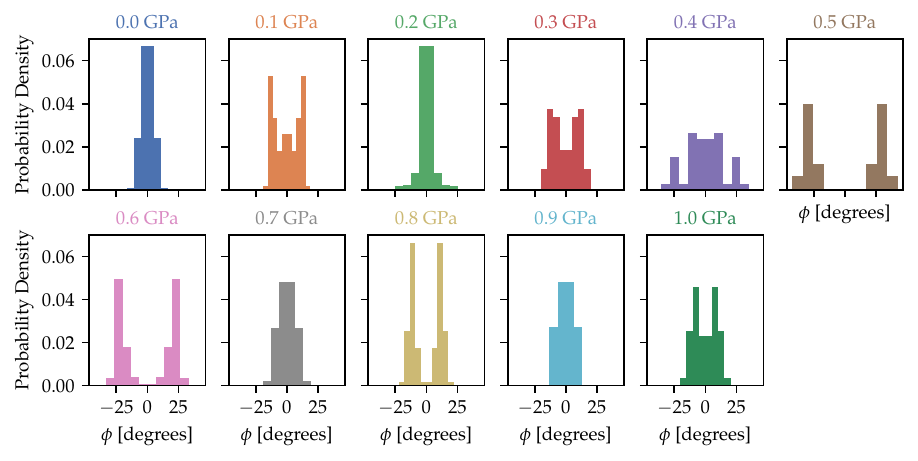}
    \caption{Probability densities of the dihedral angle in a single
    quench from with target temperature 300 K and at several fixed pressures. Here, as in Fig.~\ref{fig:histograms_280K}, a single quench simulation at each pressure is not sufficient to provide the statistics needed to show unimodal behavior.}
    \label{fig:histograms_300K}
\end{figure}

\begin{figure}[H]
    \centering   
    \includegraphics[width=\textwidth]{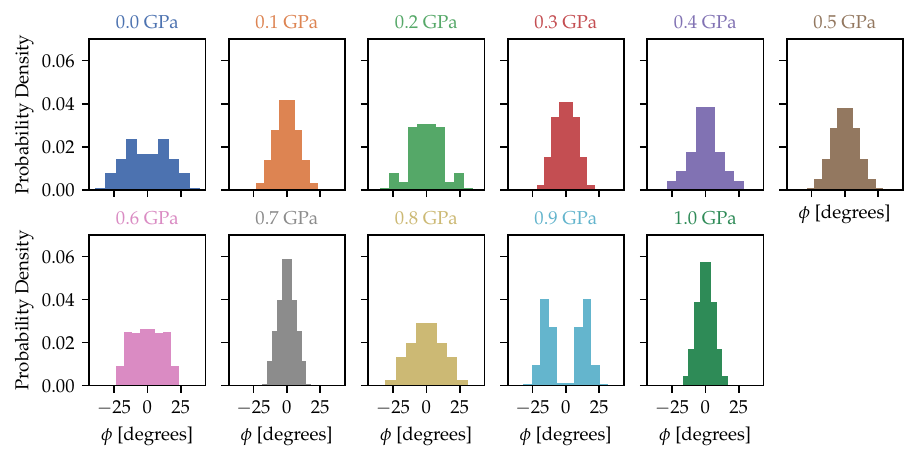}
    \caption{Probability densities of the dihedral angle in a single
    quench from with target temperature 320 K and at several fixed pressures. In this case, owing to the higher temperature, a single quench simulation at each pressure is sufficient to provide the statistics needed to show unimodal behavior, for most pressures considered here.}
    \label{fig:histograms_320K}
\end{figure}

The simulations reported here are based on equilibrating the liquid at constant temperature 400 K and constant pressure for 1 ns in order to obtain a racemic mixture, quenching the system instantaneously to the desired target temperature, and maintaining the latter temperature constant for 500 ns. Throughout this process the pressure is kept constant. Below, we call this simulation protocol \textit{quench simulations}. The target temperatures considered here are 230, 280, 300, and 320 K and the resulting probability densities of the order parameter (average dihedral angle) are shown in Figures~\ref{fig:histograms_230K} to~\ref{fig:histograms_320K}, respectively. These temperatures are well below the glass transition for the model considered here and thus it is not possible to equilibrate the liquid, even in the relatively long simulations times that we employed.

We provide a short description and analysis of the result in each figure's caption. In short, at all target temperatures the probability densities of the order parameter are unimodal, if sufficient statistics can be obtained either by performing many simulations or by using higher temperatures for which the liquid is close to equilibrium. Therefore, these results bolster our conclusion regarding the absence of chiral symmetry breaking in our model.

\newpage
\section*{Cumulative mean square displacement vs. temperature}

We analyzed the cumulative mean square displacement (MSD) as a function of temperature to evaluate the variation of this property across the glass transition. The cumulative MSD is $\langle \left | \mathbf{r}(t)-\mathbf{r}(0) \right |^2\rangle$, where $\mathbf{r}(t)$ are atomic coordinates at time $t$ and the average $\langle \cdot \rangle$ is taken over all particles in the system. 

\begin{figure}[H]
    \centering   
    \includegraphics[width=\textwidth]{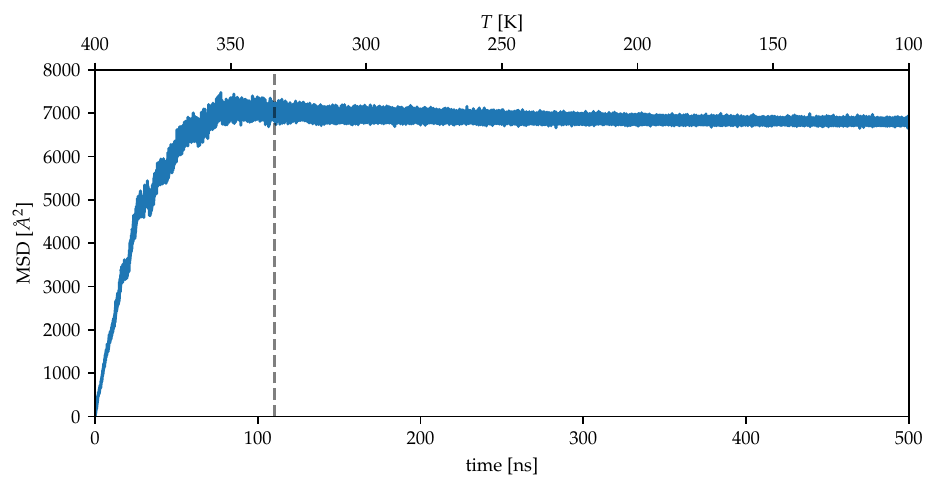}
    \caption{Cumulative mean square displacement (MSD) vs. time in a cooling simulation at a pressure of 1 atm $\approx$ 0.1013 MPa. The top axis shows the simulation temperature and the vertical line is the glass transition temperature calculated using the method described in the main part of the article.}
    \label{fig:MSD}
\end{figure}

We calculated the cumulative MSD during a cooling simulation from 400 K to 100 K at a constant pressure of 1 atm $\approx$ 0.1013 MPa. The results are shown in Fig.~\ref{fig:MSD} and we observe that the slope of the curve diminishes upon cooling and at around 350 K the curve is already flat, signaling saturation of the cumulative mean squared displacement and hence a significantly reduced atomic diffusion. In the same figure, we show the glass transition temperature calculated using the method described in the main part of the article and the agreement is satisfactory.

\end{document}